# Notes on "An Effective ECC based User Access Control Scheme with Attribute based Encryption for WSN"

Mrudula S
M.Tech Ist Year
NBKRIST – Nellore
A.P, India
mrudula.s911@gmail.com

M.ChandraMouli Reddy
PhD Scholar
Veltech Univ
Chennai-600062, T.N
mouli.veltech@gmail.com

Lakshmi Narayana K
Asst prof-CSE
SITAMS
Chittoor-A.P
kodavali.lakshmi@gmail.com

JayaPrakash. P
Asst prof-CSE
SITAMS
Chitoor-A.P
pokalajayaprakash@gmail.com

Chandra Sekhar Vorugunti
DA-IICT
Gandhinagar-382007
Gujarat
vorugunti_chandra_sekhar@daiict.ac.in

*Abstract*— **The rapid growth of networking and communication technologies results in amalgamation of 'Internet of Things' and 'Wireless sensor networks' to form WSNIT. WSNIT facilitates the WSN to connect dynamically to Internet and exchange the data with the external world. The critical data stored in sensor nodes related to patient health, environment can be accessed by attackers via insecure internet. To counterattack this, there is a demand for data integrity and controlled data access by incorporating a highly secure and light weight authentication schemes. In this context, Santanu et al had proposed an attribute based authentication framework for WSN and discussed on its security strengths. In this paper, we do a thorough analysis on Santanu et al scheme; to show that their scheme is susceptible to privileged insider attack and node capture attack. We also demonstrate that Santanu et al scheme consists of major inconsistencies which restrict the protocol execution.**

## I. INTRODUCTION

The advances in internet technologies resulted in a dynamic internet called "Internet of Things" which can be labelled as a worldwide interconnection of distinctively addressable objects like RFID tags, sensors, Owens etc. The amalgamation of WSN and IoT allows mutual communication between external world and WSN by exchanging the information (patient health readings, environment data etc) sensed by sensors via Internet. However, accessing the sensor node via Internet raises security challenges, which need to be addressed to gain the advantage of the various benefits of such combination.

To achieve the data integrity and controlled access to sensor data, there is a demand for secure authentication schemes to allow only the legitimate user having specified access attributes to connect to the WSN. Due to the resource constrained nature of sensor nodes, the authentication scheme should not result in execution of heavy weight cryptographic operations like encryption, decryption etc. by the sensor nodes.

Various researchers had proposed protocols for secure authentication of users connecting to WSN [1-12] based on various techniques like password based [1-12], Temporal Credential based [1], biometric based [1,7], chinese remainder theorem based [3,4], identity based [5], bilinear pairing [6], ECC based [5,6,9,10], chaotic map based [8], attribute based [10,11,12] etc.,. Unfortunately, most of the protocols are analyzed insecure shortly, after they were put forward [1,2,3,4,5,6,8].

In 2015, Santanu et al. [10] proposed an ECC-based user access control scheme with attribute-based encryption for WSN and claimed that their protocol achieves stronger security by resisting major cryptographic attacks. In this paper we will show that Santanu et al. scheme is completely vulnerable to privileged insider attack which leads to leakage of user password to an insider, vulnerable to node capture attack, which leads to leakage of user identity to an attacker. Also Santanu et al scheme requires huge data storage and computation cost for generating user smart card. We will also show that, Santanu et al scheme consists of many inconsistencies or anomalies in various phases of their protocol execution.

The rest of the paper is organized as follows. In section II, a brief review of Santanu et al. scheme is given. Section III, describes the security weakness of Santanu et al. scheme. In section IV, we discuss various inconsistencies of Santanu et al. scheme, section V provides the conclusion of the paper.

## II. REVIEW OF SANTANU ET AL SCHEME

In this section, we examine an effective ECC-based user access control scheme with attribute based encryption for WSN by Santanu et al [10] in 2015. The notations used in Santanu et al. [10] are listed below.

$U_i$: User
BS: Base station
$S_j$: Sensor node
$CH_u$: Cluster head ' u'
$PW_i$: Password of user $U_i$
$ID_i$: Identity of user $U_i$
$CID_u$: Identifier of cluster head $CH_u$
$h(.)$: A secure one-way hash function.
$x$: Secret Key of base station
$X$: Public Key of base station.
$MKCH_u$: Master key for cluster head $CH_u$

**Registration Phase:**

     $U_i$                                    **Base Station (B.S)**

Selects: $ID_i$, $r_i$, and $PW_i$.
Computes: $RPW_i = h(ID_i \| r_i \| PW_i)$.

                $\{RPW_i, ID_i\}$

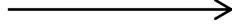

                                     Receive at time $TSU_i$
                                     Choose '$y_i$' specific to $U_i$
                                     Compute: $A_i = h(ID_i \| TSU_i)$
                                            $R_i = h(RPW_i \| A_i)$
                                     $U_i$ public key $Y_i = y_i.G$
                                     Selects an access structure $P_i$ for each user $U_i$.
                          For each node 'x' in $P_i$, the B.S needs to construct a $d_x+1$ degree
polynomial $q_x$ applying the Lagrange interpolation, where $d_x$ is the degree of the node 'x'.
The BS then selects 'm' deployed cluster heads in the network $CH_1, CH_2, .. CH_m$,
Computes: m key-plus-id combinations $\{(S_{iu}, CID_u) | 1 <= u <= m\}$, where $S_{iu} = h(TSU_i \| (T_{CHu} \oplus W_{CHu}))$.

S.C = {:(i) G (ii) $TSU_j$ (iii) $P_j$ (iv) $RU_i$ (v) $K_{iu}$ for each leaf node u $\varepsilon$ $P_i$, (vi) $Y_i$, (vii) h(.), and
        (viii) m+m' key-plus-id combinations $\{(S_{iu}, CID_u) | 1 <= u <= m+m'\}, r_i\}$ .

---

**Login Phase :**

     $U_i$                                    **Smart Card (S.C)**

provides $ID_i$, $PW_i$.
                               $RPW_i^* = h(ID_i \| r_i \| PW_i)$, $A_i^* = h(ID_i \| TSU_j)$
                               $R_i^* = h(RPW_i^* \| A_i^*)$ // check computed $R_i^*$ equals to the $R_i$ stored in the S.C.
                               Select a session specific arbitrary number $n_i$.
                               Computes $N_i = n_i.G$. $SK_i = n_i.Y_i$,
                               $M_i = h(R_i^* \| TU_i \| N_{ix})$, where $N_{ix}$ is the x-co-ordinate of $N_i$,
                               $TU_i$ is the current time stamp.
                               Selects a cluster head, say $CH_u$
                               $T_{iu} = TU_i \oplus S_{iu}$.
                               $<N_i \| E_{SKix}(ID_i \| CID_u \| T_{iu} \| M_i)>$ to the BS, via a public channel.

---

**Authentication Phase :**

**Base Station (B.S)**                               **Cluster Head 'U' ($CH_u$)**

Receive : $<N_i \| E_{SKix}(ID_i \| CID_u \| T_{iu} \| M_i)>$ at $TSU_i$
Computes: $SK_i^* = y_i.N_i$.
'$y_i$' is a secret key assigned to $U_i$ by B.S during registration phase.
Decrypts $D_{SKix}(E_{SKix}(ID_i \| CID_u \| T_{iu} \| M_i)$ to get $\{ID_i, CID_u, T_{iu}, M_i\}$.
$S_{iu}^* = h(TSU_i \| (T_{CHu} \oplus W_{CHu}))$,
$TSU_i$ is the time stamp at which the B.S received the $U_i$ registration request.
$TU_i^* = T_{iu} \oplus S_{iu}^*$, $M_i^* = h(R_i \| TU_i \| N_{ix})$
Check $M_i^*$ = received $M_i$. If both are equal, B.S confirms the authenticity of $U_i$.
BS computes $T1 = h(TU_i^* \| T_{BS})$
Computes: $E_{MKCHu}(ID_i \| CID_u \| T_{iu} \| T_{BS} \| N_i \| T1 \| Y_i \| TSU_i)$.

                 $<E_{MKCHu}(ID_i \| CID_u \| T_{iu} \| T_{BS} \| N_i \| T1 \| Y_i \| TSU_i)>$

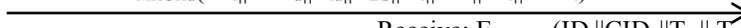

                               Receive: $E_{MKCHu}(ID_i \| CID_u \| T_{iu} \| T_{BS} \| N_i \| T1 \| Y_i \| TSU_i)$
                               $D_{MKCHu}[E_{MKCHu}(ID_i \| CID_u \| T_{iu} \| T_{BS} \| N_i \| T1 \| Y_i \| TSU_i)]$
                               Validate $CID_u$.
                               Check if $|T_{BS} - T_{BS}^*| < \Delta T$, where $T_{BS}^*$ is the received time.
                               Computes : $S_{iu}^* = h(TSU_i \| (T_{CHu} \oplus W_{CHu}))$ // stored in $CH_u$ memory

$$TU_i^* = T_{iu} \oplus T_{CHu}, \quad T1^* = h(TU_i^* \| T_{BS}).$$
Check: computed $T1^*$ equals the received $T1$

*Master Key Encryption Phase :*

*Cluster Head 'U' (CHu)*          *User ($U_i$)*

Compute: $K3 = (B_u + Y) \pmod{p}$
$B_u$ and $Y$ are stored in CHu memory
$K_{2u} = (B_u + T_u)$ for all the attributes $u \varepsilon I_u$
$K_{iu} = h(CIDu \| ID_i \| TU_i^* \| N_{ix} \| TCHu)$, $K_{si} = h(K_{iu} \| K3)$
$M1 = E_{Ksi}(M)$.

$<M1 \| TS_{CHu} \| (T_{CHu} \oplus W_{CHu}) \| T_{CHu} \| h(ID_i \| CIDu \| TS_{CHu} \| M1) \| K_{2u}$ for all $u \varepsilon I_u >$ $\longrightarrow$

*Data Decryption Phase*

*User ($U_i$)*          *Cluster Head 'U' (CHu)*

Receive :
$<M1 \| TS_{CHu} \| (T_{CHu} \oplus W_{CHu}) \| T_{CHu} \| h(ID_i \| CIDu \| TS_{CHu} \| M1) \| K_{2u}$ for all $u \varepsilon I_u >$
Compute: $h(ID_i \| CIDu \| TS_{CHu} \| M1)$
Compare the computed hash with the received.
Check $|TS_{CHu}^* - TS_{CHu}| < \Delta T$
Retrieve $W_{CHu}$ from $T_{CHu} \oplus W_{CHu}$
Validate $W_{CHu}$.
Access$(K_u U_i + K_{2u}) = ((q_x(0) - t_i)g + (B_i + T_i)) \pmod{p}$
$= (q_r(0) - t_i)g + (b_i + t_i)g) \pmod{p} =$
$(y + b_i)g \pmod{p} = (y_g + b_i g) \pmod{p} = K3^*$
where $q_x(0) = q_{parent(x)}^{(index(x))}$, $q_r(0) = y$ and $q_x(0) = q_{parent(x)}^{(index(x))}$
$K_{iu} = h(CIDu \| ID_i \| TU_i \| N_{ix} \| TS_{CHu})$ and $K_{si} = h(K_{iu} \| K3^*))$
$D_{Ksi}(M1) = D_{Ksi}(E_{Ksi}(M)) = M$

*A. Pre Deployment Phase*

The base station (BS) performs the following steps in offline mode before the actual deployment of the sensor nodes and cluster heads in deployment field.

**Step 1.** B.S selects a large odd prime number 'p' of minimum 160 bits, generates a Galois Field G.F (p) and elliptic curve $E_p(a,b)$, which is a set of all points on the curve $y^2 = x^3 + ax + b \pmod{p}$, such that $a, b \varepsilon Z_p = \{0, 1, 2, 3 .... p-1\}$, satisfying the condition $4a^3 + 27b^2 \neq 0$. 'G' represents the base point of elliptic curve 'E' of order 'n', which is of 160 bits such that $n > \sqrt{p}$. B.S chooses a random number 'x' as its secret private key, computes its public key $= X = x.G$.

**Step 2.** B.S generates a random number $t_i$ for each attribute $i \varepsilon$ universe of all the sensor attributes set I. Computes $T_i = t_i.G$.

**Step 3.** Finally, the BS loads the following information into the memory of each cluster head CHu (u = 1, 2, . . ., m): a unique identifier CIDu, assigns a set of attributes $I_u$ from global attribute set I. Assigns a master key $MK_{CHu}$ and a secret key $K_u$ and $B_u = b_u.G$ ($b_u$ is known only to B.S ), a time stamp $T_{CHu}$ and its expiration time $W_{CHu}$. B.S pre loads CIDu, $E_P(a,b)$, G, $I_u$, $MK_{CHu}$, $B_u$, $K_u$, $T_u$, $T_{CHu}$, $W_{CHu}$.

**Step 4.** For each deployed sensor node $S_j$: The B.S assigns a unique identity $SID_j$ and assigns a set of attributes $I_j$ from global set I. Assigns a master key $MKS_j$. B.S pre loads $SID_j$, $I_j$, $MKS_j$ for each sensor node.

*B. Registration Phase*

This phase is invoked whenever a user $U_i$ registers with the base station for the first time.

**Step 1.** The user $U_i$ selects the identifier $ID_i$, a random number $r_i$, and the password $PW_i$. $U_i$ then computes $RPW_i = h(ID_i \| r_i \| PW_i)$. $U_i$ provides the computed masked password $RPW_i$ and $ID_i$ to the base station via a secure channel for registration.

**Step 2.** On receiving the login request $\{RPW_i, ID_i\}$ at time $TSU_i$, the BS computes the following variables for $U_i$. $A_i = h(ID_i \| TSU_i)$, secret masked information $R_i = h(RPW_i \| A_i)$. B.S also selects a secret key '$y_i$' for each user $U_i$ and computes the $U_i$ public key $Y_i = y_i.G$.

**Step 3.** The B.S selects an access structure $P_i$ for each user $U_i$. After receiving the registration information from the valid users, the B.S assigns the access structure $P_i$ for each $U_i$. These access structures are implemented through access tree. Every leaf nodes of the access tree is labeled with an attribute like 'Doctor' or 'Nurse' etc. The non-leaf nodes are reflected as a threshold gates. The access structures are symbolized using the logic expressions over the attributes. The privilege of user $U_i$ is defined with the help of an access tree. For each node 'x' in $P_i$, the B.S needs to construct a $d_x+1$ degree polynomial $q_x$ applying the Lagrange interpolation, where $d_x$ is the degree of the node 'x'.

**Step 4.** For each non-root node 'x' in $P_i$, it sets $q_x^{(0)} = q_{parent(x)}^{(index(x))}$, in which parent(x) is the parent of 'x' and index(x) denotes the index(x)-th child element of root node 'x'. The B.S then assigns $q_r(0) = y$, where $q_r(0)$ is the polynomial of the root of the access tree of the user $U_i$. Finally the B.S computes $K_{iu} = (q_i(0)-t_i) \pmod p$ for each leaf node $i \in P_i$.

**Step 5.** The BS then selects 'm' deployed cluster heads in the network $CH_1, CH_2, \ldots, CH_m$, which will be deployed during the initial deployment phase, and computes the m key-plus-id combinations $\{(S_{iu}, CID_u) | 1 \le u \le m\}$, where $S_{iu} = h(TSU_i || (T_{CHu} \oplus W_{CHu}))$.

**Step 6.** For dynamic cluster head addition phase, the m' cluster heads, $CH_{m+1}, CH_{m+2}, \ldots, CH_{m+m'}$, will be deployed later, after the initial deployment in the network, in order to replace some compromised cluster heads, if any, and add some fresh cluster heads along with sensor nodes. For this purpose, the BS assigns the unique identity $CID_u^*$ and unique master key $MK_{CHu}^*$. Similarly the B.S assigns the unique identity $SID_j^*$, and unique master key $MKS_j^*$.

**Step 7.** Finally, the BS issues a tamper-proof smart card with the following parameters stored in it :(i)G (ii) $TSU_i$ (iii) $P_i$ (iv)$RU_i$ (v) $K_{iu}$ for each leaf node $u \in P_i$, (vi) $Y_i$, (vii) h(.), and (viii) m+m' key-plus-id combinations $\{(S_{iu}, CID_u) | 1 \le u \le m+m'\}$. The value of m + m' is chosen according to memory availability of the smart card.

**Step 8.** On receiving the S.C from B.S, $U_i$ inserts $r_i$ into its S.C.

*C. Login Phase*

Whenever the user $U_i$ wants to access real-time data from a sensor of a deployed WSN, the user $U_i$ needs to perform the following steps:

**Step 1.** $U_i$ inserts his/her smart card into the card reader of a specific terminal and provides his/her Identity $ID_i$, password $PW_i$.

**Step 2.** The smart card then computes the masked password of the user $U_i$ as $RPW_i^* = h(ID_i || r_i || PW_i)$, $A_i^* = h(ID_i || TSU_i)$, using the time stamp $TSU_i$ stored in $U_i$ S.C. S.C further computes $R_i^* = h(RPW_i^* || A_i^*)$, and then checks whether the computed $R_i^*$ equals to the $R_i$ stored in the S.C. If this verification fails, it means $U_i$ provided the incorrect credentials and terminates the session, else the smart card proceeds to perform the following steps.

**Step 3.** The S.C selects a session specific arbitrary number $n_i$ and computes $N_i = n_i.G$. The S.C computes $SK_i = n_i.Y_i$, $M_i = h(R_i^* || TU_i || N_{ix})$, where $N_{ix}$ is the x-co-ordinate of $N_i$, $TU_i$ is the current time stamp of $U_i$.

**Step 4.** The user $U_i$ selects a cluster head, say CHu from which the real time data can be accessed inside WSN. Corresponding to CHu, the smart card computes $T_{iu} = TU_i \oplus S_{iu}$. Finally, the user $U_i$ sends the message $<N_i || E_{SKix} (ID_i || CID_u || T_{iu} || M_i)>$ to the BS, via a public channel.

*D. Authentication Phase*

On receiving the login request message $<N_i || E_{SKix} (ID_i || CID_u || T_{iu} || M_i)>$ from the user $U_i$, the BS performs the following steps in order to authenticate the user $U_i$.

**Step 1.** The BS computes $SK_i^* = y_i.N_i$ where '$y_i$' is a secret key assigned to $U_i$ by B.S during registration phase. Using this computed key $SK_i^*$, the BS decrypts $D_{SKix}(E_{SKix} (ID_i || CID_u || T_{iu} || M_i))$ to get $\{ID_i, CID_u, T_{iu}, M_i\}$.

**Step 2.** The BS computes $S_{iu}^* = h(TSU_i || (T_{CHu} \oplus W_{CHu}))$, where $TSU_i$ is the time stamp at which the B.S received the $U_i$ registration request, $T_{CHu}$ is the boot straped time stamp for CHu and $W_{CHu}$ is its validity. B.S retrives $TU_i^* = T_{iu} \oplus S_{iu}^*$ and computes $M_i^* = h(R_i || TU_i || N_{ix})$. B.S compares the computed $M_i^*$ with the received $M_i$. If both are equal, B.S confirms the authenticity of $U_i$.

**Step 3.** Using the current system time stamp $T_{BS}$, the BS computes $T1 = h(TU_i^* || T_{BS})$ and produces a cipher text message, encrypted using the master key $MK_{CHu}$ of the cluster head CHu as $E_{MKCHu}(ID_i || CID_u || T_{iu} || T_{BS} || N_i || T1 || Y_i || TSU_i)$. The BS directs the message $<E_{MKCHu}(ID_i || CID_u || T_{iu} || T_{BS} || N_i || T1 || Y_i || TSU_i)>$ to the corresponding cluster head CHu.

**Step 4.** After receiving the message in Step3 from the BS, the cluster head CHu decrypts $E_{MKCHu}(ID_i || CID_u || T_{iu} || T_{BS} || N_i || T1 || Y_i || TSU_i)$ using its own master key $MK_{CHu}$ as $D_{MKCHu}[E_{MKCHu}(ID_i || CID_u || T_{iu} || T_{BS} || N_i || T1 || Y_i || TSU_i)] = (ID_i || CID_u || T_{iu} || T_{BS} || N_i || T1 || Y_i || TSU_i)$. CHu then checks if retrieved $CID_u$ is equal to its identity i.e $CID_u$. If both are equal, CHu further checks if $|T_{BS} - T_{BS}^*| < \Delta T$, where $T_{BS}^*$ is the received system time stamp of the CHu and $\Delta T$ is the allowed valid transmission delay. If it holds good, CHu computes $S_{iu}^* = h(TSU_i || (T_{CHu} \oplus W_{CHu}))$ using its boot straped $T_{CHu}$ and its expiration time $W_{CHu}$. CHu proceeds to compute $TU_i^* = T_{iu} \oplus T_{CHu}$, using the retrieved $TU_i^*$, CHu computes $T1^* = h(TU_i^* || T_{BS})$. If the computed $T1^*$ equals the received T1, CHu authenticates B.S. Finally, CHu performs master key encryption phase which forwards the login reply message to $U_i$. $U_i$ performs the data decryption phase in respond to master encryption phase.

*E. Master Key Encryption Phase*

In this phase, any user CHu performs the following steps:

**Step 1.** CHu computes $K3 = (B_u + Y) \pmod p$, based on the stored values of $B_u$ and Y. CHu further computes $K_{2u} = (B_u + T_u)$ for all the attributes u, where $u \in I_u$ for that cluster head CHu.

**Step 2.** As per the user $U_i$ request, the cluster head $CH_u$ computes $K_{iu} = h(CID_u\|ID_i\| TU_i^*\| N_{ix}\|TC_{Hu})$, $K_{si} = h(K_{iu}\| K3)$. $CH_u$ encrypts the sensor data M using $K_{si}$ i.e $M1 = E_{Ksi}(M)$.

**Step 3.** $CH_u$ forwards the authentication reply message $<M1\|TS_{CHu}\|(T_{CHu} \oplus W_{CHu})\| T_{CHu}\|h(ID_i\|CID_u\| TS_{CHu}\| M1)\| K_{2u}$ for all u ε $I_u$ > to $U_i$, where $TS_{CHu}$ is the current time stamp of $CH_u$.

### F. Data Decryption Phase

**Step 1.** On receiving the message from the cluster head $CH_u$, during the master key and data encryption phase, the user $U_i$, computes a hash value $h(ID_i\|CID_u\|TS_{CHu}\| M1)$ based on the received values of $TS_{CHu}$, M1. If the computed hash value equals the received hash value, $U_i$ proceeds further, else terminates the session. $U_i$ further compares $|TS_{CHu}^* - TS_{CHu}| < \Delta T$, if it is valid, $U_i$ further computes $W_{CHu}$ from $T_{CHu} \oplus W_{CHu}$ and validates the expiration time i.e $W_{CHu} >= T_{CHu}$. If the condition is valid, $U_i$ authenticates $CH_u$ else terminates the connection.

**Step 2.** On successful authentication of $CH_u$, $U_i$ starts the decrypts process from leaf nodes of its own access tree, in a bottom to top up approach. $U_i$ computes $F_a$ for each leaf node 'l' in $P_i$ using the following logic: if 'a' ε $I_u$, the $F_a$ = Access $(K_uU_i + K_{2u})$ else $F_a = \frac{1}{2}$ (invalid). For the user $U_i$, the access tree is $P_i$ for which the root node is 'r'. If $P_{ix}$ is the sub-tree of $P_i$, $P_{ix}$ is rooted at node x. If a set of attributes $I_u$ satisfies the access tree $P_{ix}$, then only we obtain Access $(K_uU_i^*g + K_{2i}) = (y+b_i)g \pmod{p}$, which is shown in the following steps: Access$(K_uU_i+ K_{2u}) = ((q_x(0)-t_i)g+(B_i+T_i)) \pmod{p} = (q_r(0)-t_i)g+(b_i+t_i)g) \pmod{p} = (y+b_i)g \pmod{p} = (y_g+b_ig) \pmod{p} = K3^*$ where $q_x(0) = q_{parent(x)}^{(index(x))}$, $q_r(0) = y$ and $q_x(0) = q_{parent(x)}^{(index(x))}$, is executed in a recursive way as discussed in [12].

**Step 3.** The user $U_i$ computes $K_{iu} = h(CID_u\|ID_i\|TU_i\| N_{ix}\| TS_{CHu})$ and $K_{si} = h(K_{iu}\|K3^*))$ and then uses the key $K_{si}$ to decrypt $D_{Ksi}(M1) = D_{Ksi}(E_{Ksi}(M)) = M$.

### III. CRYPTANALYSIS OF SANTANU ET AL SCHEME

In this segment, we will cryptanalyze the Santanu et al [10] scheme and illustrate that Santanu et al scheme is vulnerable to privileged insider attacker, node capture attack. Also we will illustrate that Santanu et al scheme consists of few anomalies or inconsistencies which restricts the protocol flow.

*Attack Model:*

1) An opponent or an attacker or legal user can extract the information cached in the smart card by several techniques such as power consumption or leaked information [10] etc. i.e S.C = {:(i) G (ii) $TSU_j$ (iii) $P_j$ (iv) $RU_i$ (v) $K_{iu}$ for each leaf node u ε $P_i$, (vi) $Y_i$, (vii) h(.), and (viii) m+m' key-plus-id combinations {( $S_{iu}$, $CID_u$) | 1<= u <= m+m'}, $r_i$ }.

2) An opponent can passive monitor or eavesdrop or alter or replay the login request, login reply messages communicated among $U_i$, B.S, CH over a public channel which is Internet. i.e  {{$N_i\|E_{SKix}(ID_i\|CID_u\|T_{iu}\|M_i$}, { $E_{MKCHu}(ID_i\|CID_u\|T_{iu}\|T_{BS}\|N_i\|T1\|Y_i\|TSU_i)$}, {<M1\| $TS_{CHu}$ \| ($T_{CHu} \oplus W_{CHu}$)\| $T_{CHu}\|h(ID_i\| CID_u\|TS_{CHu}\| M1$) \| $K_{2u}$ for all u ε $I_u$ >}}

### A. Huge Data Storage and Computation Requirement for Generating User Smart Card

In Santanu et al. scheme the user smart card memory is stored with 'm' key-plus-id combinations {($S_{iu}$, $CID_u$)| 1<=u <= m}, where $S_{iu} = h(TSU_i\|(T_{CHu} \oplus W_{CHu}))$ of all cluster heads in the WSN. Based on the Santanu et al. discussion, for a total of 20,000 sensor nodes to be deployed and if each cluster head can handle 200 sensor nodes, then there are total m = 100 cluster heads needed and for dynamic cluster head addition m' =100 cluster heads are reserved. So a total of m+m' = 100+100 ($S_{iu}$, $CID_u$) details are stored. A total of 200 hash operations need to be performed for each user smart card. If the system contains 'n' users, then a total of (n * 200) hash operations need to be performed to load the smart card memory of corresponding user which requires huge computation cost from the BS. The major issue is that, the user may not interested or in need of data from all the cluster heads. Hence storing all the m+m' cluster head details is a major drawback in Santanu et al scheme.

Suppose, if any cluster head $CH_u^*$ is found to be compromised, the B.S re assigns the master key $MK_{CHu}^*$, identity $CID_u^*$, set of identities $I_u^*$, $B_u^* = b_u.G$ ($b_u$ is known only to B.S), a time stamp $T_{CHu}^*$ and its expiration time $W_{CHu}^*$. Apart from re assigning the values of compromised $CH_u^*$, the S.C values of all the users whose key-plus-Id combination contains $CH_u$ must be modified, which requires huge computation cost.

### B. Fails to Resists Privileged Insider Attack

As demonstrated by A.K Das et al in their recent work [1], (one of the authors of the Santanu et al [10] scheme), performed an insider attack, we will apply same strategy on Santanu et al scheme, which is described as follows:

**Step 1.** Assume that the privileged insider 'E' stolen or gets the smart card of $U_i$ for a while, then, 'E' can launch offline password guessing attack as follows:

**Step 2.** As discussed in 2.3, during registration phase of $U_i$, $U_i$ submits the login request {$RPW_i$, $ID_i$} where $RPW_i = h(ID_i\| r_i \|PW_i)$. From the login request 'E' intercepts $ID_i$, from the stolen smart card of $U_i$, 'E' can get $r_i$. Now from the intercepted value of $RPW_i = h(ID_i\| r_i \|PW_i)$, 'E' can perform guessing attack [1] on $U_i$ password from a moderately small dictionary, over the subsequent steps.

**Step 2.1).** Choose a guessed password $PW_i$ to be $PW_i^*$.
**Step 2.2).** Compute $RPW_i^* = h(ID_i\|r_i\|PW_i^*)$ and compare the computed $RPW_i^*$ with the intercepted value of $RPW_i$ during registration phase. If $RPW_i^*$ equals $RPW_i$, then the guessed password $PW_i^*$ is the correct one, else 'E' proceeds with another guessed password $PW_i^*$ from the dictionary and follows the same steps.

Hence, in Santanu et al scheme, the insider on getting the stolen S.C of $U_i$ can compute the $U_i$ password.

*C. Fails to Resists Node Capture Attack*

A sensor node capture attack is a WSN specific physical type of attack. As the sensor nodes and cluster heads are not equipped with tamper resistant hardware, the attacker can capture a legitimate sensor node or cluster head in WSN and can extract the secret information stored in it. Each cluster head and sensor node is pre-loaded with the master keys i.e $MK_{CHu}$ and $M.K.S_j$ respectively. On receiving the login request $<N_i||E_{SKix}(ID_i||CIDu||T_{iu}||M_i)>$ from $U_i$, the base station computes and forwards a message to the corresponding cluster head $CH_u$ which is encrypted with its master key i.e $<E_{MKCHu}(ID_i||CID_u||T_{iu}||T_{BS}||N_i||T1||Y_i||TSU_i)>$. If an attacker captures the cluster head $CH_u$ can retrieve its master key $MK_{CHu}$ which is stored in its hardware, as discussed above, an attacker 'E' can capture all the communications between the protocol entities, hence, 'E' can capture the message sent by B.S to $CH_u$ and can decrypt it to get $D_{MKCHu}(E_{MKCHu}(ID_i||CID_u||T_{iu}||T_{BS}||N_i||T1||Y_i||TSU_i)) = \{ID_i, CIDu, T_{iu}, T_{BS}, N_i, T1, Y_i, TSU_i\}$ which involves the identity of the logged in user i.e $ID_i$. Therefore, we can conclude that, in Santanu et al scheme, if an attacker captures a cluster head, he can extract the secret key stored in it and can decrypt the messages forwarded to the cluster head to get the identity of the user logged in. Hence, we can conclude that Santanu et al scheme fails to resist identity leakage attack.

IV. ANALYSIS OF INCONSISTENCIES IN SANTANU ET AL SCHEME

In this section, we discuss few anomalies or inconsistencies found in Santanu et al [10] scheme.

*A. Inconsistencies in Registration Phase:*

In Santanu et al scheme, during registration phase, the Base Station B.S computes the variables $A_i = h(ID_i||TSU_i)$, $R_i = h(RPW_i||A_i)$ where $TSU_i$ is the time at which the B.S received the registration request from $U_i$. B.S also selects a secret key $y_i$ for $U_i$ and computes the $U_i$ public key $Y_i = y_i.G$.

*1) Inconsistency 1:*

In Santanu et al scheme, the authors didn't mentioned clearly on how the base station stores the value '$R_i$' and the secret key '$y_i$' assigned to each user in its database, to use these values in authentication phase.

*B. Inconsistencies in Authentication Phase:*

In this section, we discuss the inconsistencies found in authentication phase of Santanu et al scheme.

*1) Inconsistency 1:*

In Santanu et al scheme, during the registration phase, S.C submits the login request message $<N_i||E_{SKix}(ID_i||CIDu||T_{iu}||M_i)>$ to the BS, in which $N_i = n_i.G$, where '$n_i$' is an arbitrary number chosen by S.C specific to the current login request. $SK_i = n_i.Y_i$, where $Y_i = y_i.G$ is the $U_i$ public key computed out of a secret key '$y_i$' assigned to $U_i$ by the B.S. $SK_{ix}$ is the x-co-ordinate of $SK_i$.

On receiving the login request from $U_i$, B.S proceeds to compute $SK_i = y_i.N_i$. To compute $SK_i$, B.S needs '$y_i$', which is secret key of $U_i$ assigned by B.S. The login request contains two parts $N_i$, which is a random one and doesn't contains any clue about the user identity. The second part of login request $E_{SKix}(ID_i||CIDu||T_{iu}||M_i)$ is encrypted using $SK_{ix}$, which is also a random value and doesn't contain any clue about the user details. Therefore it is impossible for a B.S to identify from which user, it got the login request. Hence, it is impossible to retrieve '$y_i$' a user specific value to compute $SK_i$.

*2) Inconsistency 2*

In Santanu et al scheme, as discussed in section 2.5 (authentication phase), step 2, the B.S computes $M_i^* = h(R_i ||TU_i|| N_{ix})$. To compute $M_i^*$, B.S requires the value '$R_i$'. In Santanu et al scheme, the B.S is computing '$R_i$' during $U_i$ registration phase, but as discussed in 4.1.2, it is not clear in which format, the B.S stores $R_i$ in its data base. One option for B.S is to compute $R_i$ i.e $R_i^* = h(RPW_i^*||A_i^*)$ where $A_i^* = h(ID_i||TSU_i)$, $RPW_i = h(ID_i || r_i ||PW_i)$. B.S doesn't know $U_i$ password $PW_i$, identity $ID_i$, random number chosen by $U_i$ i.e $r_i$. Hence, it is impossible to compute $R_i^*$ by B.S. In summary, B.S is not storing $R_i^*$ and doesn't have the required parameters to compute $R_i^*$. Therefore, computing $M_i^* = h(R_i ||TU_i|| N_{ix})$ is impossible for B.S.

V. CONCLUSION

Recently Santanu et al. proposed an ECC-based user access control scheme with attribute-based encryption for WSN. Even though it is a novel attempt, after thorough analysis of Santanu et al paper, we demonstrated that their scheme is vulnerable to privilege insider attack and node capture attack. We also established that Santanu et al scheme include major inconsistencies which oppose the correct protocol execution. In future work, we aim to propose a secure and light weight authentication scheme for WSN by eliminating the security pitfalls and inconsistencies found in Santanu et al scheme.